\documentclass[useAMS,usenatbib,usegraphicx]{mn2e}

\title{The life-time of galactic bars: \\central mass concentrations and gravity torques}
\author[F. Bournaud, F. Combes and B. Semelin]{F. Bournaud$^{1}$\thanks{E-mail: Frederic.Bournaud@obspm.fr}, F. Combes$^{1}$, and B. Semelin$^{1}$\\
$^{1}$Observatoire de Paris, LERMA, 61 av. de l'Observatoire, F-75014, Paris, France}
\begin{document}

\date{Accepted XXXX. Received XXXX; in original form XXXX}

\pagerange{\pageref{firstpage}--\pageref{lastpage}} \pubyear{2002}

\maketitle

\label{firstpage}

\begin{abstract}
Bars in gas-rich spiral galaxies are short-lived. They 
drive gas inflows through their gravity torques, and at the same time 
self-regulate their strength. Their robustness has been subject of debate, since it was thought that only the resulting central mass concentrations (CMCs) were weakening bars, and only relatively rare massive CMCs were able to completely destroy them. Through numerical simulations including gas dynamics, we find that with the gas parameters of normal spiral galaxies, the CMC is not sufficient to fully dissolve the bar. But another overlooked mechanism, the transfer of angular momentum from the infalling gas to the stellar bar, can also strongly weaken the bar. In addition, we show that gravity torques are correctly reproduced in simulations, and conclude that bars are transient features, with life-time of 1-2 Gyr in typical Sb-Sc galaxies, because of the combined effects of CMCs and gravity torques, while most existing works had focussed on the CMC effects alone.
\end{abstract}


\section{Introduction}
In most barred galaxies, gas is concentrated on the leading side of the bar \citep[e.g.,][]{dvc63}, which is reproduced in numerical simulations \citep[e.g.,][]{AT92}. Then, the bar gravity torques make gas lose angular momentum, which initiates a gas inflow and fuels a central mass concentration (CMC). The bar is weakened by the CMC growth, because of escaping orbits \citep{Hasan90, Pfenniger90}. \citet{Friedli93}, \citet{Berentzen98} and \citet{Hozumi99} report the dissolution of galactic bars with CMCs of mass 0.5 to 2 \% of the disk mass. To explain the large fraction of barred galaxies observed both at $z\simeq 0$ \citep[e.g.,][]{Esk02} and at $z>0.7$ \citep{sheth}, we have proposed that bars are dissolved and reformed \citep[ hereafter paper I]{BC02}: our N-body simulations showed that bars are dissolved in 1--4 Gyrs in most galaxies, which we had attributed to the CMC growth. Yet, whether bars are really transient is still debated. Resonant rings are often observed in barred galaxies, which implies that gravitational torques are much larger than viscous torques \citep[e.g.,][]{Buta96}. For a long time, gaseous simulations were too limited in resolution, induced too large viscous torques, so that such rings did not form. If viscous torques are still over-estimated, this may induce unrealistic inflows of gas, so that CMCs are more massive and/or fueled more rapidly, and the life-time of bars may be much under-estimated \citep{Regan04}. Moreover, even if the CMC fueling is realistic, the effects of the CMC may not be enough to fully dissolve the bar: recent simulations with more spatial resolution by \citet{shen03} -- that do not include the whole gas response -- have shown that bars are more robust against the growth of CMCs than what was believed before. 

Thus, the limited lifetime of bars found in paper~I and other works could 
be an artifact of viscous torques or an over-estimation of the effects of 
the CMC. In this Letter, we show that the gas inflow in our simulations is 
actually initiated by gravitational torques, that are much larger than 
viscous torques. While most existing works had focused on the effects of 
the CMC growth, we show that another phenomenon can also 
lead to the destruction of galactic bars in a few Gyrs: the torques 
between the stellar bar and the gaseous arms also disturb the bar and can dissolve it. We then find that galactic bars are transient, even when the gas infall and the effects of the CMC are correctly treated and do not destroy the bar themselves: the growth of the CMC is not the only factor of bar dissolution.

%==========================================================

\section{Numerical simulations}
We employ the N-body FFT code described in paper I, including star formation. The gravitational potential is computed on a Cartesian grid of size $512\times 512\times 64$. The softening length and cell size are 75 pc, and the number of particles is $10^6$ for each component (gas and stars).

 The dissipative dynamics of the ISM is modeled by the 3-D sticky-particles code described in paper I (we use an elasticity parameter for cloud-cloud collisions $\beta_r=\beta_t=0.75$ in this paper). Star formation and stellar mass-loss are also included. We use a visible (stellar and gaseous) mass of $2\times 10^{11}$ M$_{\sun}$. The disk truncation radius is $14$ kpc and its scale-height is 1.2~kpc. The bulge and dark halo are Plummer spheres of radial scale-lengths 2 and 36 kpc respectively. The bulge-to-disk mass ratio is 0.25, and the dark-to-visible ratio inside the disk radius is 0.6. The initial gas mass is $14.5\times10^9$~M$_{\sun}$ (7.25 \% of the visible mass). At $t=1.5$ Gyr, during the bar dissolution, the remaining mass of gas is $11.3\times 10^9$  M$_{\sun}$ (5.5 \% of the visible mass). We also run a control simulation without gas. 
%In Sect.~4, we use a second simulation with a bulge-to-disk mass ratio of 0.33, and an initial gaseous mass of $9\times 10^9$ M$_{\sun}$ (4.5 \% of the visible mass). 
We compute the bar strength $P_2$ as the maximum over radius of the $m=2$ angular component of the gravitational torques, following the definition given in paper I, where this parameter was denoted $S_2$ (see eq.~27 in paper~I).

%=======================================================================

\section{Mechanisms for bar dissolution}
\subsection{Gas inflows and bar dissolution}

\begin{figure}
\centering
\includegraphics[width=55mm]{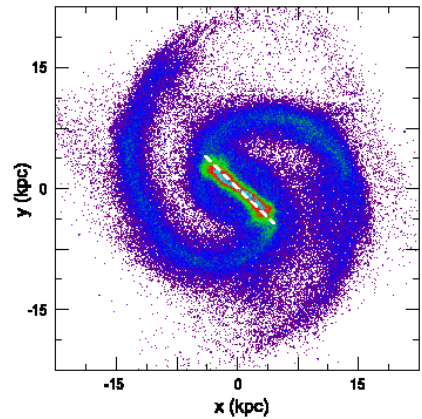}
\caption{Face-on plot of the gas density at $t=800$ Myr. The dashed line represents the bar major axis, defined as the major axis of the most eccentric ellipse in an ellipse-fitting model of the stellar density. Gas is concentrated on the leading side of the bar inside the CR radius (6 kpc), with a shift of several degrees.}\label{snap}
\end{figure}

\begin{figure}
\centering
\includegraphics[width=75mm]{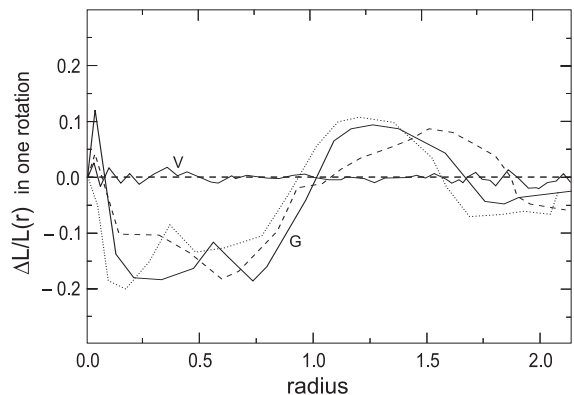}
\caption{Angular momentum lost by gas in one rotation, in units of the inital angular momentum $L(r)$, as a function of radius (in units of the CR radius at $t=900$ Myr). The curves labeled ''G'' show the effects of the gravitational torques; they are mean values over periods of 100 Myr around $t=800$ (dotted line), 900 (solid line), and 1000 Myr (dashed line). The curve labeled ''V'' shows the effects of viscous torques, that are negligible with respect to gravity torques. Inside the CR radius, gas loses 10--15~\% of its momentum in one period. Torques are positive between the CR and the outer Lindblad resonance (OLR). Because of the bar slowing-down, the CR and OLR radii increase with time. The thin solid line shows the comparison with an SPH simulation with the same initial conditions at $t=900$~Myr. Gravity torques are of the same order with this different code.
}\label{fig2}
\end{figure}

The response of gas to the stellar bar is shown in Fig.~\ref{snap}. Inside the corotation radius (CR, 6kpc), gas is concentrated on the leading side of the bar. Then, as shown in Fig.~\ref{fig2}, the bar gravity torques make gas lose 10--15\% of its angular momentum over one rotation, while the shear viscosity torques are much smaller: the gas inflow in our simulations is really initiated by the bar gravity torques, contrary to the hypothesis raised by \citet{Regan04}.

The bar-driven gas inflow fuels a central mass concentration (CMC): we show in Fig.~\ref{fig1} the increase in the mass inside radii 100 and 250 pc. At $t=2$ Gyr, the mass radial profile shows a central peak of stars and gas that can be fitted by a Plummer CMC of mass $2.8\times 10^9$ M$_{\sun}$ and radial scale-length 90 pc. According to \citet{shen03}, such a CMC cannot fully dissolve the bar. Yet, while the gravitational softening length and the number of particles we used are close to that of \citet{shen03}, the bar is here fully dissolved in less than 2 Gyrs: the residual values of $P_2\simeq 0.15$ correspond to a weak oval distortion and spiral arms, but no bar is present.

A surprising point shown in Fig.~\ref{fig1} is that the CMC growth occurs after the bar is already significantly weakened. The increase of the mass included inside radius 250 pc occurs between $t=1.2$ and $t=1.8$ Gyr. The bar strength is then smaller than 40\% of its maximal value. As for the increase in the mass inside radius 100 pc, it occurs at the very end of the bar dissolution, from $t=1.4$ to $t=1.8$ Gyrs. At the opposite, the bar weakening is already significant at $t=1$ Gyr: at this time, no strong increase in the mass has occurred, excepted at large scales of more than 1kpc, but mass concentrations at such scales are totally inefficient to dissolve bars. It thus seems that the CMC growth is not the only factor that is responsible for the destruction of the bar. Furthermore, the bar dissolution is not an artifact of a too long time step: we have divided the time step by factor 8, and found a similar result (see Fig.~\ref{fig1}).

%=======================================================================

%\section{Mechanisms for bar dissolution}

\subsection{Effects of the CMC}

\begin{figure}
\centering
\includegraphics[width=70mm]{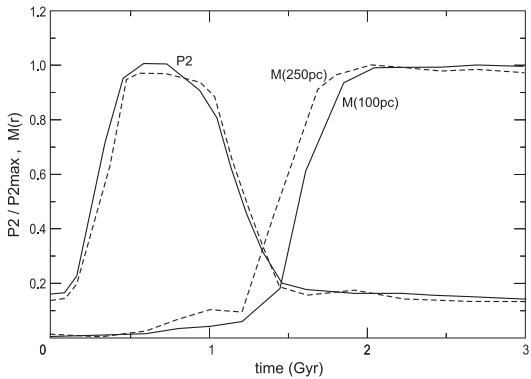}
\caption{Evolution of the bar strength in the main simulation (solid line labeled ''P2'') and in the simulation with an eigth times shorter time step (dashed), and increase in the central masses inside radii 250pc (dashed line) and 100 pc (solid line). The initial masses have been subtracted, so that these curves only show the bar-induced growth of a CMC. All the quantities have been rescaled to 1 at their maximum value.
}\label{fig1}
\end{figure}

\begin{figure}
\centering
\includegraphics[width=70mm]{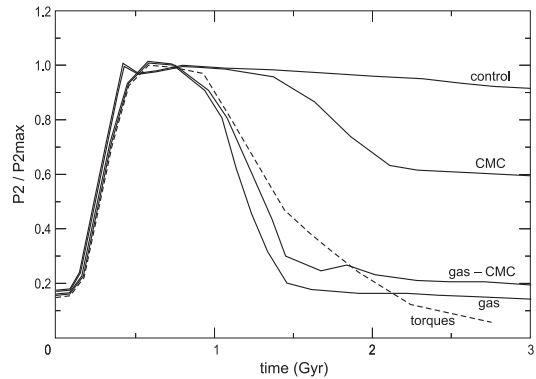}
\caption{
Evolution of the bar strength in several models: 
''gas'': complete simulation including gas -- 
''control'': stellar simulation without gas -- 
''CMC'': stellar simulation without gas and addition of a growing CMC --
''gas-CMC'': complete simulation including gas and artificial removal of the CMC growth --
''torques'': stellar simulation without gas and addition of gravity torques representing those exerted by gas on the stellar bar.
%Evolution of the bar strength in the complete simulation (green, labeled 'gas'), the purely stellar simulation (red, labeled 'control'). The blue curve labeled 'CMC' shows the result of the stellar model in which the final CMC is analytically added. The orange curve corresponds to the simulation including gas in which the CMC is artificially suppressed by removing the particles that have fallen inside the 90 central pc. The dashed black curve labeled 'torques' shows the result of the stellar simulation, in which the gravitational torques exerted by gas on the stellar bar are analytically added. these different models show that the CMC is cannot completely dissolve the bar; on the contrary, the bar is mainly destroyed by the gravity torques.
}\label{P2a}
\end{figure}

To confirm that the CMC growth is not the only phenomenon responsible for the bar dissolution, we have run a purely stellar simulation with the same initial conditions, and added an analytical CMC. This model is similar to that used by \citet{shen03}. The final CMC is a Plummer sphere corresponding to the CMC we had fitted in the complete simulation (see previous Section). Its radial scale-length is 90 pc, and its mass is grown up to $2.8\times 10^9$ M$_{\sun}$ following the variations of M(100pc) shown in Fig.~\ref{fig1}. In this model, the bar is not fully dissolved, but only partly weakened (see Fig.~\ref{P2a}). This is in agreement with the conclusion of \citet{shen03} that the bar can survive to the growth of such a CMC.

We have also run a simulation in which we include the gas response, but 
suppress artificially the growth of the CMC: every gas particle that is 
found below radius 100 pc, with a velocity inferior to the circular velocity at this radius, is artificially suppressed, unless already present at $t=0$. We do not suppress particles that are just crossing the central 100 pc with a large velocity, but only those that have definitively fallen in this region. As shown in Fig.~\ref{P2a}, the bar is dissolved, and the evolution is very similar to that of the complete simulation. This result definitively proves that a phenomenon different from the CMC growth intervenes in the dissolution of the bar. This phenomenon is related to the gas response, since it is not observed in the purely stellar simulation (Fig.~\ref{P2a}).

\subsection{Effects of gravity torques}

Inside the CR, the bar has been shown to exert important gravitational torques on the ISM. In the same time, the leading arms of gas exert torques on the stellar bar. The mean torque undergone by stars inside the bar is positive, but the real torques vary with radius and azimuth. This could be the reason why the bar is dissolved in the simulation where the CMC has been artificially suppressed: gravitational torques between the stellar bar and gaseous arms are still present in this model.

To confirm the role of gravity torques, we have run a purely stellar simulation in which gravitational torques exerted by gas have been artificially added. To this aim, in the complete simulation, we compute the amplitude $A_i$ and phases $\phi_i$ of the tangential forces $F_\mathrm{T}^\mathrm{gas}$ exerted by gas in the Fourier decomposition 
\begin{equation}
F_\mathrm{T}^\mathrm{gas} (r,\theta)= \sum_i A_i(r) \sin\left (i (\theta - \phi_i \right))
\end{equation}
where $\theta$ is computed in the frame of the bar. We record the mean values $<A_i>$ and $<\phi_i>$ between 0.8 and 1 Gyr. Then, in a purely stellar simulation, we analytically add the gravity torques given by
\begin{equation}
F_\mathrm{T}^\mathrm{gas} (r,\theta)= \sum_{i=2,4,6,8} <A_i>(r) \sin\left (i (\theta - <\phi_i> \right))
\end{equation}
where $\theta$ in still computed in the frame of the stellar bar. The amplitude of these torques is realistic, since $<A_i>$ and $<\phi_i>$ are given by the complete simulation, even if they do not follow any temporal evolution. We apply no torques before 800 Myr and increase them progressively up to their final amplitude between 800 and 900 Myr. In this model, that only differs from the purely stellar simulation by the addition of these gravitational torques, the bar is fully dissolved (see Fig.~\ref{P2a}), and its dissolution is nearly as rapid as in the complete simulation. 

\begin{figure}
\centering
\includegraphics[width=70mm]{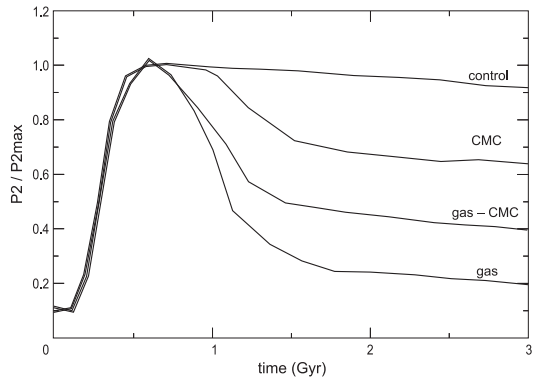}
\caption{
Same as Fig.~\ref{P2a} with an initial gas-to-visible mass ratio of 4.5\% and a bulge-to-disk mass ratio of 0.33. The bar is dissolved in the complete simulation (curve labeled ''gas''). The effect of gravity torques (curve labeled ''gas-CMC'') participate much to the bar weakening, but the CMC also plays a role.
}\label{P2b}
\end{figure}

Then, the gravity torques exerted by infalling gas on the bar can strongly weaken it. In this first simulation, this process dissolves it, while the role of the CMC is negligible. We have run another simulation with an initial gaseous mass equal to 4.5 \% of the visible mass, and a bulge-to-disk mass ratio of 0.33, and made the same models with the CMC alone and the torques alone. The results are shown in Fig.~\ref{P2b}. In the complete simulation, the bar is fully dissolved, while the growth of the CMC only partly weakens it. Here again, the gravitational torques play a major role in the bar dissolution. They are not strong enough to fully dissolve the bar alone, but the combined effects of the CMC lead to a complete dissolution. This second case confirms the role of gravity torques in the dissolution of bars, even is the CMC plays a role, too.

%======================================================================

\section{Discussion and conclusion}
 
In our simulations, viscous torques are negligible with respect to gravitational torques and do not directly initiate the gas inflow. Yet, the phase shift between the bar and gas, and the magnitude of the resulting gravity torques, are largely dependent on the viscosity \citep{alabda81}. Then, whether gravity torques in our simulations are realistic or not has to be checked.

\begin{figure}
\centering
\includegraphics[width=70mm]{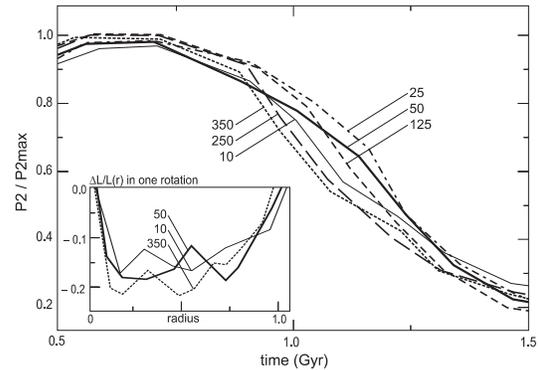}
\caption{Evolution of the bar strengh with different box sizes in the sticky-particle scheme, for the first simulation detailed before. The thick solid line represent this main simulation (see also Fig.~1), with a box size of 50 pc. Other lines represent box sizes from 10 pc to 350 pc. The bar dissolution is not largely affected by this parameter. The inset shows the gravity torques for collision box sizes of 10, 50 and 350 pc (see Fig.~2 for details). The torques are slightly larger when the box size is larger, but the change in gravity torques is rather small: it does not exceed the variations observed in 100 Myr (see Fig.~2).
}\label{visco}
\end{figure}

\begin{figure}
\centering
\includegraphics[width=70mm]{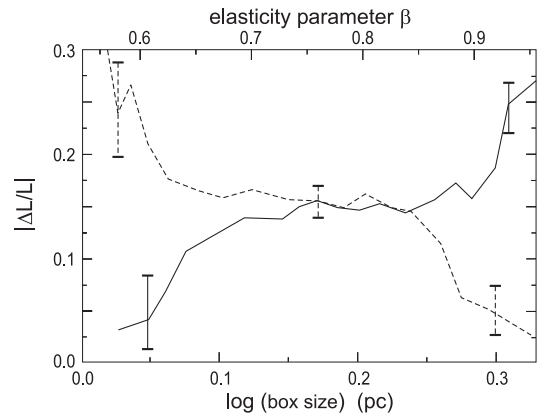}
\caption{Variations of the gravity torques exerted on the gas, as a function of the box size (solid curve, bottom axis, with an elasticity parameter $\beta=0.75$), and as a function of the elasticity parameter (dashed curve, top axis, with a box size of 50 pc). the quantity $\Delta L/L$, plotted in absolute value, is defined as in Sect.~3.1. Gravity torques are not strongly affected by these parameters in the range 10--400pc for the box size and $\beta=$0.65--0.85 for the elasticity parameter. This explains why the bar evolution is not strongly affected by a change in these parameters, as shown in Fig.~6. Outside this particular range, torques are more affected by these two parameters, but the ISM is either unrealistically cold or unrealistically stable : only parameters in the range where gravity torques are not strongly affected are acceptable.}\label{fig_params}
\end{figure}

First, we have checked that our results are not dependent on the box size used to compute particles collisions in the sticky-particles scheme: this parameter controls the distance along which particles interact, influences the numerical viscosity, thus the gravity torques may depend on it. For box sizes ranging from 10 pc to 350 pc, we find that changes in this parameter does not lead to large changes in the gravity torques, and the bar life-time is not dependent on this parameter, as shown on Fig.~\ref{visco}. It may seem surprising that changing this box size parameter does not influence the bar life-time at all. In fact, as shown by Fig.~\ref{fig_params}, the gravity torques are actually dependant on the sticky particles parameters (the box size, and the elasticity parameter $\beta$ for cloud-cloud collisions): gravity torques increase with the box size, and decreases with the elasticity parameter. But there is a range of parameters where gravity torques are nearly constant, for elasticity parameter $\beta$ between 0.65 and 0.85, and box sizes between 10pc and 350~pc. For larger box sizes and/or smaller $\beta$, gravity torques become larger, but the medium is very cold (with a typical sound speed smaller than 5~km~s$^{-1}$) and clumpy, instead of forming spiral arms: these situations are not realistic, so they are ruled out. At the opposite, large $\beta$ or small box sizes lead to situations with smaller gravity torques, but the dissipation rate in the ISM is not realistic, the sound speed in the ISM is found to be larger than 10 or even 15~km~s$^{-1}$ and the Toomre parameter for gas is larger than 1.5: as a result, gas only forms very smooth spiral arms that disappear after a few dynamical times. Thus, the only acceptable parameters are $0.65<\beta<0.85$ and box sizes from 10 to 350 pc (for our number of gas particles). In this range, there is an equilibrium between dissipation in the ISM and kinematical heating by density waves: for instance a smaller $\beta$ leads to more dissipation, so the first spiral arms are stronger and heat the medium more. This explains why physical properties are not strongly affected by the gas dynamical parameters in this particular range. Then, provided that we choose parameters resulting in a realistic model for the ISM (not too cold or too hot, i.e. a medium in which the main instabilities are spiral arms), the physical results are not strongly affected by the sticky particles parameters, so they are robust.

\begin{figure}
\centering
\resizebox{6.6cm}{!}{\includegraphics{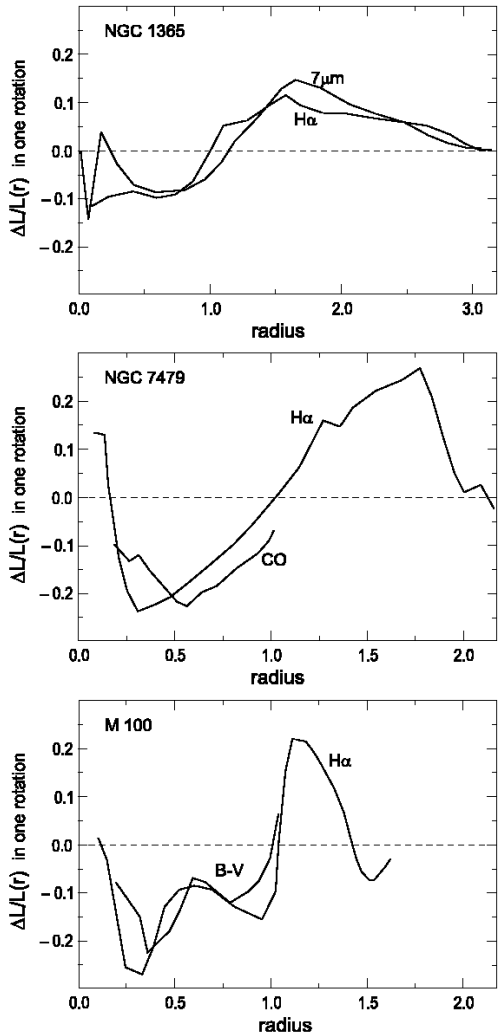}}
\caption{
Gravity torques exerted by the bar on the gaseous arms in NGC~1365, NGC~7479 and M~100, as function of the radius (in units of the CR radius), for several tracers of the ISM (NGC~1365: H$\alpha$ and dust emission at 7$\mu$m -- NGC~7479: CO and H$\alpha$ -- M~100: H$\alpha$ and dust absorption traced by the B-V index). The gravitational potential of bars is estimated from I-band images. The CR radius has been defined as the radius where the averaged torques become positive. Gas inside the CR loses 10 to 20\% of its momentum over one period: this is fully compatible with our simulations, which proves that the gravity torques between the bar and gas are realistic in our simulations.
}\label{7479}
\end{figure}

As a second verification, we have estimated the gravity torques exerted by stellar bars on gaseous arms in a sample of observed barred galaxies (see a forthcoming paper for details). We show on Fig.~\ref{7479} the results for three well-known barred galaxies, NGC~1365, NGC~7479, and M~100: for these galaxies as for other barred ones that we have studied, inside the CR, gas loses typically 10--20\% of its angular momentum in one rotation. This result is independant of the tracer of interstellar gas that we choose (see Fig.~\ref{7479}). This is compatible with our simulations (Fig.~2). Thus, the gravity torques are correctly reproduced in our sticky-particles simulations, that provide a realistic estimation of the life-time of bars.

Finally, we also ran a simulation with the same initial conditions as our first run, with the tree-SPH code by \citet{SC02}, using $2\times 10^5$ particles, isothermal initial conditions at $10^4$~K, and adiabatic evolution. As shown in Fig.~2, the gravity torques between the stellar bar and gas have the same order of value as in sticky-particle simulations, and in observations.

While the dissolution of bars had always been attributed to the growth of CMCs, we find here that another process contributes to their destruction. Even if growing CMCs do not completely dissolve most galactic bars \citep{shen03}, the gravitational torques exerted by gas on the bar can strongly weaken it: these torques are positive, which make orbits become rounder; they also induce a phase shift that depends on the radius, which misaligns stellar orbits, so that the bar is strongly weakened. In the two simulations with parameters typical of Sb-Sc galaxies, the cycle of bar dissolution takes about 2 Gyrs (for the mass and radius of the Milky-Way). The supply of angular momentum from the ISM to the bar, which dissolves it, occurs much faster than the removal of momentum by the darrk halo, that could trigger the bar over timescales of several Gyrs \citep{AT02}. The constant bar fraction up to $z\simeq 1$ \citep{sheth, jogee, ee} cannot then be interpreted in terms of robust, long-lived bars. As suggested by \citet{Sell99}, the bar could be renewed either by galaxy interactions, or by accretion of external gas. \citet{Ber04} have shown that galaxy interactions can reform bar only in gas poor galaxies, i.e. not in most spirals. This indicates that gas accretion plays a major role in the high fraction of barred galaxies, which can account in detail for observed properties of bars \citep{block}.

%======================================================================

\section*{Acknowledgments}
The simulations were computed on the Fujitsu NEC-SX5 of the CNRS computing center, at IDRIS. We are grateful to the anonymous referee for useful comments.

\end{document}